\def\D0{D\O}  \def\d0{D\O}
\documentclass[12pt]{iopart}
\usepackage{epsfig}
\usepackage{iopams}  

\def\ib#1,#2,#3{       {\it ibid.\/ }{\bf #1} (19#2) #3}
\def\ap#1,#2,#3{       {\it Ann.~Phys.~(NY)\/ }{\bf #1} (19#2) #3}
\def\ijmp#1,#2,#3{     {\it Int.\ J.~Mod.\ Phys.\/ } {\bf A#1} (19#2) #3}
\def\mpl#1,#2,#3 {     {\it Mod.~Phys.~Lett.\/ } {\bf A#1} (19#2) #3}
\def\npb#1,#2,#3{       {\it Nucl.\ Phys.\/ }{\bf B#1} (19#2) #3}
\def\npps#1,#2,#3{     {\it Nucl.\ Phys.~B (Proc.~Suppl.)\/ }{\bf B#1}
                             (19#2) #3}
\def\plb#1,#2,#3{      {\it Phys.\ Lett.\/ }{\bf B#1} (19#2) #3}
\def\pr#1,#2,#3{       {\it Phys.\ Rev.\/ }{\bf #1} (19#2) #3}
\def\prd#1,#2,#3{      {\it Phys.\ Rev.\/ }{\bf D#1} (19#2) #3}
\def\prep#1,#2,#3{     {\it Phys.\ Rep.\/ }{\bf #1} (19#2) #3}
\def\prl#1,#2,#3{      {\it Phys.\ Rev.\ Lett.\/ }{\bf #1} (19#2) #3}
\def\pro#1,#2,#3{      {\it Prog.~Theor.\ Phys.\/ }{\bf #1} (19#2) #3}
\def\rmp#1,#2,#3{      {\it Rev.~Mod.~Phys.\/ }{\bf #1} (19#2) #3}
\def\sp#1,#2,#3{       {\it Sov.~Phys.~Usp.\/ }{\bf #1} (19#2) #3}
\def\zpc#1,#2,#3{      {\it Z.~Phys.\/ }{\bf C#1} (19#2) #3}
\def\epjc#1,#2,#3{      {\it Eur.\ Phys.\ J.\/ }{\bf C#1} (19#2) #3}
\def\appb#1,#2,#3{     {\it Acta Phys.\ Polon.\/ }{\bf B#1} (19#2) #3}
\def\etal{\it et al.}
\def\ie{{\it i.e.\ }}
\begin{document}

\rightline{CPPM-P-1999-01}
\rightline{IFT/98-20}
\rightline{OUNP-99-02}
\rightline{MZ-TH/99-01}

\title[High $P_{T}$ Leptons and $W$ Production]{High $P_{T}$ Leptons and
  $W$ Production at HERA}

\author{C.Diaconu~$^{1}$, J.Kalinowski~$^{2}$, T.Matsushita~$^{3}$,
  H.Spiesberger~$^{4}$, D.S.Waters~$^{3}$}

\address{$^{1}$ CPPM, Universit\'e d'Aix-Marseille II, IN2P3-CNRS,
  Marseille, France} 
\address{$^{2}$ Instytut Fizyki Teoretycznej, Uniwersytet Warszawski,
  PL-00681 Warszawa, Poland} 
\address{$^{3}$ Nuclear and Astrophysics Laboratory, Keble Road, Oxford
  OX1 3RH, UK} 
\address{$^{4}$ ThEP, Institut f\"ur Physik,
  Johannes-Gutenberg-Universit\"at Mainz, D-55099 Mainz, Germany} 

\begin{abstract}
  Details are given of the observation by H1 of events containing high
  $P_{T}$ leptons in addition to large missing $P_{T}$.  A closely
  related ZEUS analysis, including a preliminary measurement of the $W$
  production cross section, is discussed and the two experiments are
  compared. Some possible non-Standard Model sources for the events are
  considered.
\end{abstract}

\section{Introduction}

The observation by the H1 experiment of a number of events containing
high $P_{T}$ leptons in addition to large missing $P_{T}$, apparently in
excess of the number expected from Standard Model processes, has aroused
much recent interest and is outlined in section~\ref{sec:h1} of this
article.  The Standard Model background is expected to be dominated by
$W$ production, a preliminary cross section for which has been measured
by the ZEUS experiment.  ZEUS have also searched for high $P_{T}$ tracks
in events with missing $P_{T}$ using similar cuts to H1, the results of
which are presented in section~\ref{sec:zeus}. A direct comparison
between the H1 and ZEUS detector acceptances is shown in
section~\ref{sec:comp}. Finally, some theoretical speculations about
possible sources of such events in the context of $R_{p}$-violating SUSY
are presented in section~\ref{sec:theory}.

\section{H1 High $P_{T}$ Lepton Events} \label{sec:h1}

The H1 analysis~\cite{h1_iso_lepton} is based on an inclusive search for
events with a transverse momentum imbalance measured in the calorimeter,
$P_{T}^{\mathrm calo}$, greater than $25$~${\rm GeV}$.  This cut
minimizes the contributions from neutral current and photoproduction
processes and has a well understood experimental efficiency. In the
selected event sample, 124 events contain high energy tracks with
transverse momentum above 10~${\rm GeV}$ and polar angles with respect
to the proton direction above $10^{\circ}$. The vast majority of these
events are charged current events containing a high $P_{T}$ track close
to the centre of an hadronic jet. The track isolation with respect to
calorimetric deposits (${\rm D_{jet}}$) and with respect to other tracks
(${\rm D_{track}}$) is quantified by the Cartesian distance in the
$\eta-\phi$ plane~\footnote{Both H1 and ZEUS coordinate systems are
  right-handed with the $Z$-axis pointing in the proton beam direction
  and the horizontal $X$-axis pointing towards the centre of HERA. The
  pseudorapidity variable $\eta$ is related to the polar angle by
  $\eta=-\ln(\tan(\theta/2))$.}.  Six events are found to contain
isolated tracks with ${\rm D_{jet}}>1.0$ and ${\rm D_{track}}>0.5$ .

Lepton identification algorithms, based on the signal shape in the
calorimeter and muon chamber hits, indicate that the six tracks in fact
correspond to high $P_{T}$ leptons~: one event contains an electron
($e^{-}$) and five events contain muons (2 $\mu^+$, 2 $\mu^-$ and one
very energetic muon corresponding to a stiff track whose sign cannot be
determined). The muon events are labelled $\mu1$ to $\mu5$ in the
following. One of the muon events ($\mu3$) also contains a positron with
a lower transverse momentum $P_{T}(e^+)=6.7$~${\rm GeV}$.

\begin{figure}[!tb]
 \begin{center}
 \epsfig{file=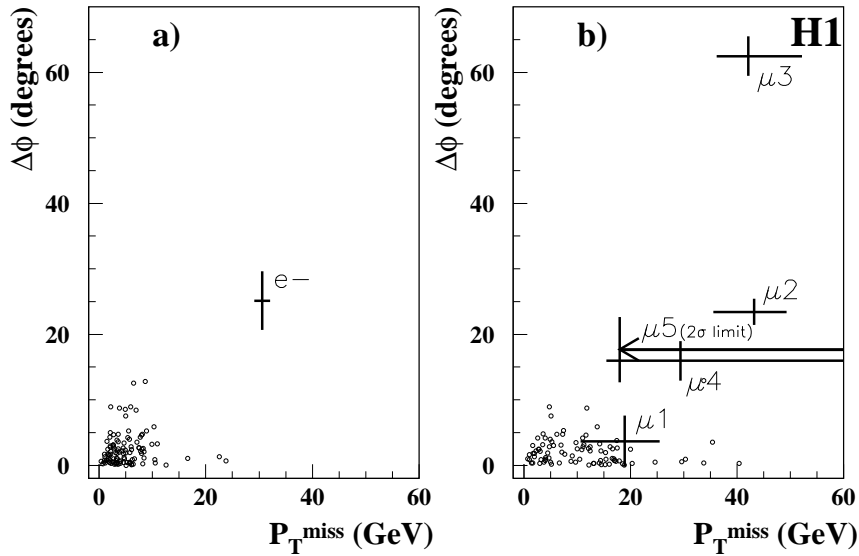,width=0.8\textwidth}
 \end{center}
\caption{Distribution of events in $P_T^{\rm miss}$ and azimuthal 
  acoplanarity $\Delta\phi$~: a)~electron channel; b)~muon channel. The
  six events are displayed with their individual measurement errors. The
  open circles show the distributions for a neutral current control
  sample, described in the text.}
\label{fig:h101}
\end{figure}

The lepton signature in each case has been investigated in detail and
found to be consistent with the assigned hypothesis. For the electron
candidate the shower pattern recorded in the calorimeter is compatible
with the expectation for an electromagnetic shower, while the isolated
track measured in the central tracker has a specific ionisation
consistent with a single particle. The muon candidates are measured in
the central tracking system, calorimeters and external iron yoke
instrumented with muon chambers. For all tracks the specific ionisation
in the central tracker is consistent with single minimum ionising
particles. The energy depositions in the calorimeters sampled over more
than 7 interactions lengths and the signals in the muon chambers are
compatible in shape and magnitude with those expected from a minimum
ionising particle. The probability that an isolated charged hadron would
simulate a muon in both the calorimeter and the instrumented iron is
estimated to be less than $3\times10^{-3}$.

In all events a hadronic shower has been detected in the calorimeters.
In the event $\mu5$ no charged particles are found in the core of the
high-$P_T$ hadronic jet. In all events an imbalance in the net
transverse momentum indicates the presence of at least one undetected
particle. This hypothesis is supported by the large value for the
lepton-hadron acoplanarity observed in most of the events, defined as
the angle in the transverse plane between the hadronic system and the
direction opposite to that of the high $P_{T}$ lepton. The significance
of the transverse momentum imbalance and acoplanarity is tested with
data using neutral current (NC) events, which are expected to be
intrinsically coplanar and balanced in $P_{T}$. For comparison to the
muon events, the kinematics in the NC sample is reconstructed using the
positron track parameters instead of calorimetric information. The six
high $P_{T}$ lepton events are compared to the NC control sample in
figure~\ref{fig:h101}.  The probability for an NC event to have both
$\Delta\phi$ and $P_T^{\rm miss}$ values greater than those measured in
a given candidate is estimated from a high statistics simulation to be
1\% for $\mu1$ and less than 0.1\% for the other candidates.

\begin{figure}[!tb]
  \begin{center}
  \mbox{\epsfig{file=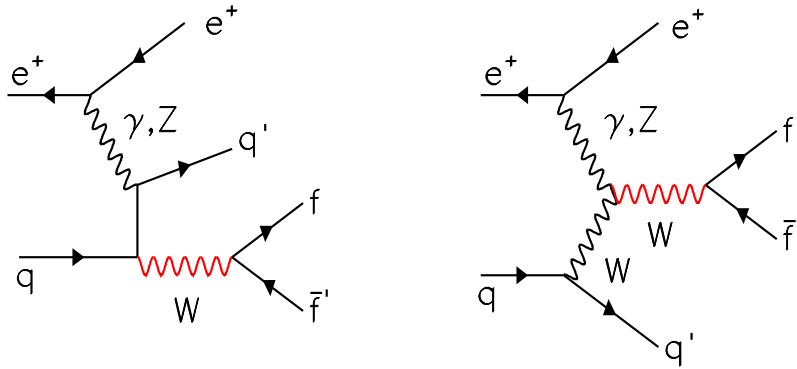,width=0.7\textwidth}}
  \end{center}
\caption{Two Feynman diagrams for the reaction $e^{+}p\to e^{+}W^{\pm}X$. The first 
  is one of the dominating diagrams, the second one involves couplings
  at the $WW\gamma$ vertex.}
\label{fig:h102}
\end{figure}

The Standard Model predictions for processes yielding events with
isolated leptons and missing energy have been investigated. The
predicted rates are dominated by $W$ production via the reaction
$e^{+}p\to e^{+}W^{\pm}X$, two diagrams for which are shown in
figure~\ref{fig:h102}, followed by the leptonic decay of the $W$. The
cross section of around $60$~${\rm fb}$ per charge state and leptonic
decay channel for this process, calculated using the program
EPVEC~\cite{BVZ_w}, gives an expected $1.7\pm0.5$ events in the electron
channel and $0.5\pm0.1$ events in the muon channel.  A recent next to
leading order calculation of the resolved photon contribution to the
cross section gives a total cross section for $e^{+}p\to e^{+}W^{\pm}X$
of $0.97$~${\rm pb}$, consistent with the leading order EPVEC
estimate~\cite{spira}.  Other significant sources of events with
isolated leptons and missing transverse momentum include neutral current
DIS in the electron channel and the $\gamma\gamma\to\mu^{+}\mu^{-}$
process in the muon channel. The total predicted rates from all Standard
Model processes are $2.4\pm0.5$~events in the $e^{\pm}$ channel
(compared with 1 event observed) and $0.8\pm0.2$~events in the muon
channel (compared with 5 events observed).

\begin{figure}[!tb]
  \begin{center}
  \epsfig{figure=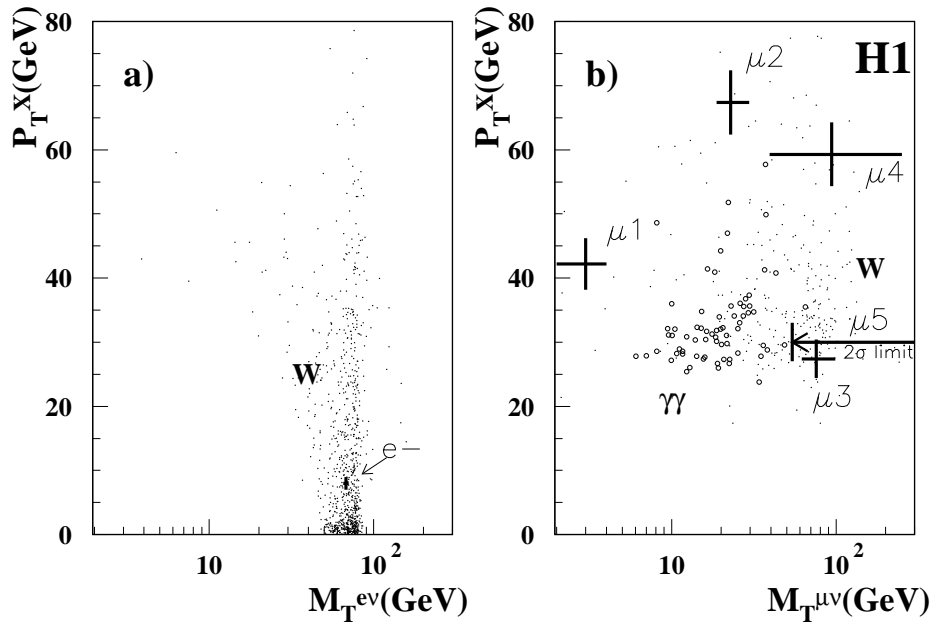,width=0.8\textwidth}
  \end{center}
\caption{Comparison of the observed events with Standard Model 
  distributions in the plane defined by the hadronic transverse
  momentum, $P_{T}^{X}$, and the lepton-neutrino transverse mass,
  $M_{T}^{\ell\nu}$. The $W$ production and
  $\gamma\gamma\to\mu^{+}\mu^{-}$ Monte Carlo samples, indicated by
  points and open circles respectively, correspond to an integrated
  luminosity a factor 500 greater than the data.}
\label{fig:h103}
\end{figure}

In figure~\ref{fig:h103} the observed events are compared to $W$
production Monte Carlo events in the plane of the transverse momentum of
the hadronic system, $P_{T}^{X}$, versus the transverse mass of the
lepton-neutrino system, $M_{T}^{\ell\nu}$. The electron event and two of
the muon events ($\mu3$ and $\mu5$) are kinematically consistent with
the Jacobian peak located around the $W$ mass and the low $P_{T}^{X}$
expected for $W$ production. Three muon events can only marginally be
accommodated within this interpretation. None of the observed muon
events are consistent with the distribution expected for
$\gamma\gamma\to\mu^{+}\mu^{-}$, also shown in figure~\ref{fig:h103}.

\section{ZEUS Results on $W$ Production and High $P_{T}$ Leptons}
\label{sec:zeus} 

The results of a search for $W$ production and leptonic decay in
$46.6$~${\rm pb^{-1}}$ of ZEUS $e^{+}p$ data have been presented
elsewhere in these proceedings~\cite{plenary_proc}. The measured cross
section from the electron channel of
$1.0$~$^{+1.0}_{-0.7}$~(stat)~$\pm$~$0.3$~(syst)~${\rm pb}$ is in good
agreement with the Standard Model prediction. The absence of any signal
in the muon channel is consistent with the smaller efficiency for
selecting events on the basis of calorimeter missing $P_{T}$, in turn a
consequence of the soft hadronic $P_{T}$ spectrum for Standard Model $W$
production.

\begin{figure}[!tb]
 \begin{center}
  \mbox{\epsfig{figure=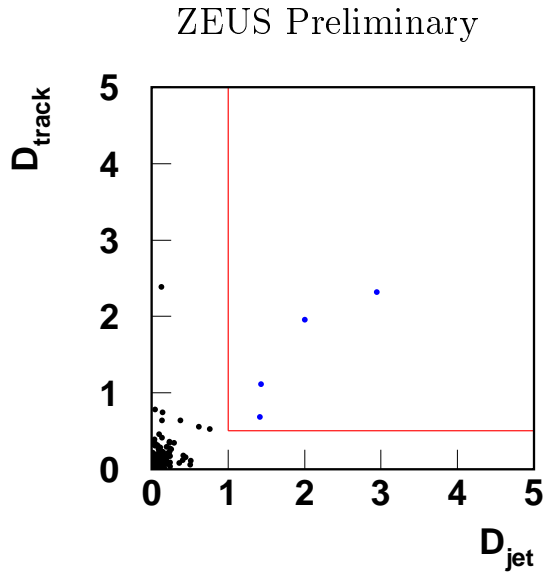}}
 \end{center}
\caption{{The jet and track $\eta-\phi$ isolation of tracks with 
    $P_{T}>10$~${\rm GeV}$ in events with calorimeter $P_{T}$ greater
    than $25$~${\rm GeV}$, for the full 1994-97 ZEUS $e^{+}p$ data.}}
\label{fig:iso_trk}
\end{figure}

In order to avoid any hidden lepton identification inefficiencies, a
separate search has been performed for isolated high $P_{T}$ vertex
fitted tracks in events with large missing $P_{T}$, applying cuts
similar to those outlined in~\cite{h1_iso_lepton}. All events with a
calorimeter $P_{T}$ greater than $25$~${\rm GeV}$ are selected, with the
exception of neutral current candidate events with an acoplanarity angle
less than $0.1$~${\rm rad}$.  The isolation variables ${\rm D_{jet}}$
and ${\rm D_{track}}$ are defined for a given track, as in the H1
analysis, as the $\eta-\phi$ separation of that track from the nearest
jet and the nearest remaining track in the event, respectively. Jets
must have $E_{T}>5$~${\rm GeV}$, an electromagnetic fraction less than
$0.9$ and an angular size greater than $0.1$~${\rm rad}$. All tracks
with $P_{T}>10$~${\rm GeV}$ in the selected events are plotted in the
$\{\rm D_{track},D_{jet}\}$ plane in figure~\ref{fig:iso_trk}. The $4$
tracks selected with ${\rm D_{jet}}>1.0$ and ${\rm D_{track}}>0.5$
agrees well with the expectation of $4.2\pm 0.6$ tracks from combined
Monte Carlo sources.
All four isolated tracks are in fact identified as positrons using
standard electron finding algorithms and criteria described
in~\cite{zeus_w}, consistent with the $2.4\pm 0.5$ ($1.5\pm 0.4$)
electron type (muon type) events expected from Monte Carlo. There is
therefore no evidence of an excess rate of high $P_{T}$ tracks, whether
identified as leptons or not, in the 1994 to 1997 ZEUS data.

\section{Comparison of H1 and ZEUS Results} \label{sec:comp}

As pointed out in~\cite{plenary_proc}, the ZEUS muon data at large
calorimeter missing $P_{T}$ disfavours high hadronic $P_{T}$ $W$
production as the source of all the H1 high $P_{T}$ muon events. This is
consistent with the kinematic properties of the H1 events themselves.
While the low statistics of the H1 and ZEUS observations cannot
currently exclude a statistical fluctuation, it is nevertheless
interesting to ask whether any source of events with a topology similar
to the H1 events would be observed at ZEUS.  In particular, the leptons
in the H1 events are concentrated at small polar angles, close to where
the ZEUS central tracking chamber track reconstruction efficiency is
expected to fall off.

\begin{figure}[!tb]
 \begin{center}
  \mbox{\epsfig{figure=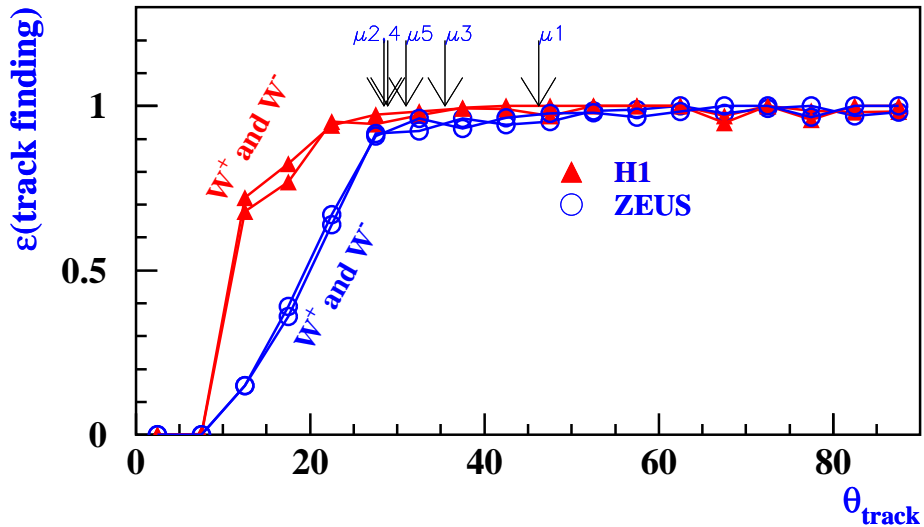,width=0.8\textwidth}}
 \end{center}
\caption{{The $P_{T}>10$~${\rm GeV}$ track finding efficiency as a 
    function of polar angle for muons from $W^{+}$ and $W^{-}$ decay
    passed through the H1 and ZEUS detector simulations. The polar
    angles of the H1 high $P_{T}$ muons are also indicated.}}
\label{fig:h1_zeus_angle_comp}
\end{figure}

Using $W$ production Monte Carlo events passed through the H1 and ZEUS
detector simulations, the efficiency with which muons from $W\to\mu\nu$
decay have a corresponding track reconstructed with $P_{T}>10$~${\rm
  GeV}$ can be calculated. The efficiencies for both $W^{+}$ and $W^{-}$
production are plotted as a function of polar angle in
figure~\ref{fig:h1_zeus_angle_comp}. Also indicated are the polar angles
of the H1 high $P_{T}$ muons, further details of which may be found
in~\cite{h1_iso_lepton}.
It can be seen that the H1 events lie in a region where the ZEUS track
reconstruction efficiency is equally high, lending weight to the
argument that a signal ought to have been seen in the ZEUS analyses
presented here. However, the positions of the H1 and ZEUS turn on curves
are significantly different and are currently being checked using
suitable data samples.

Although more data will clearly be required to fully understand the
source of the H1 high $P_{T}$ lepton events, it is nevertheless
worthwhile at this point to consider new mechanisms that might give rise
to events of this type.

\section{Theoretical Speculations} \label{sec:theory}

To date, non-Standard Model production mechanisms for the isolated muon
events have been proposed in~\cite{kkk} and~\cite{krsz}. In both papers
the discussion is performed in the framework of the supersymmetric
standard model with $R_{p}$-breaking. The primary process is the
$s$-channel production of a single scalar top quark ($\tilde{t}_1$) in
$e^+d_k$ collisions
\begin{equation}
e^+d_k\rightarrow \tilde{t} \label{rest}
\end{equation}
through the $R_{p}$-breaking interaction Lagrangian 
\begin{equation}
L=\lambda'_{13k}\cos\theta_t (\tilde{t}_1\, \bar{d}_{kR} \, e_L + 
\tilde{t}_1^* \, \bar{e}_L \, d_{kR})  
\label{lagr}
\end{equation}
where $\lambda'_{13k}$ denotes the $R_{p}$-violating coupling to the
down quark of the $k$-th generation. The angle $\theta_t$ denotes the
mixing angle in the scalar top quark sector; a similar term involving
the heavier stop $\tilde{t}_2$ is also present with $\cos\theta_t$
replaced by $\sin\theta_t$.

The interaction Lagrangian (\ref{lagr}) originates from the general
$R_p$-breaking ($\not \!\! R_p$) super\-potential
\begin{equation}
W_{\not R_p}=\lambda_{ijk}{L}_i {L}_j {E^c}_k 
+ \lambda'_{ijk}{L}_i {Q}_j {D^c}_{k} + 
\lambda''_{ijk}{U^c}_i {D^c}_j {D^c}_k, 
\label{RBW}
\end{equation}
where the left-handed lepton (quark) superfield doublets are denoted by
$L$ ($Q$), the right-handed lepton (quark) singlets by $E$ ($U$ and
$D$), and $i, j, k$ are generation indices. The first two terms violate
lepton number and the last term violates baryon number.  The couplings
$\lambda$, $\lambda'$ and $\lambda''$ are subject to many constraints
from low-energy and high-energy LEP, HERA and Tevatron data
\cite{barbier}.

The production mechanism (\ref{rest}) is based on the resonant formation
of $\tilde{t}$ in $e^+p$ collisions since the rate associated with
virtual $\tilde{t}$ production would be too small.  These phenomena
could be related to a possible surplus of high $Q^2$, high $x$ events in
neutral current scattering seen in the 1994-1996 HERA data.  However,
even if the NC events cannot be interpreted as $\tilde{t}$ resonance
production (not necessarily {\it one single} resonance), or are
interpreted as a statistical fluctuation, there is still room left for
speculation regarding the source of the isolated $\mu^+$ events in the
SUSY sector based on top squarks in the mass range of 200 -- 230~${\rm
  GeV}$, so long as the branching ratio $B_{eq}$ for the
$R_{p}$-violating decay $\tilde{t}\rightarrow e^+q$ is small and the
$R_{p}$-conserving decay modes are dominant.

\begin{figure}[htbp]
 \begin{center}
  \mbox{\epsfig{figure=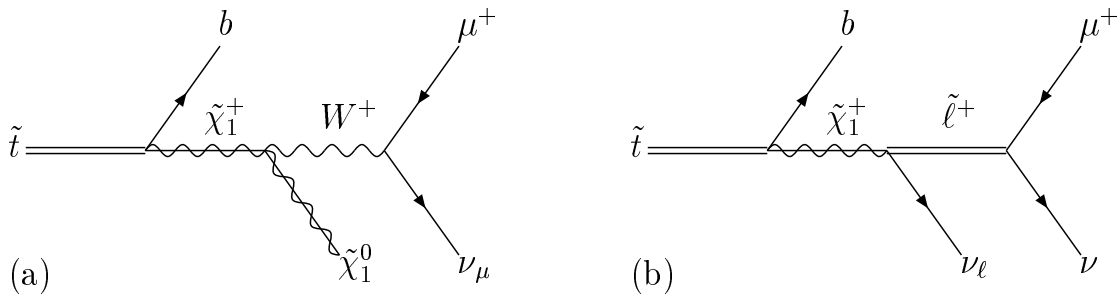,width=1.0\textwidth}}
 \end{center}
\caption{\label{decays}Possible decay chains of the stop leading to
  isolated muon + jet + missing~$P_T$: $\tilde{t} \rightarrow b
  \tilde{\chi}_1^+$ followed by (a) $\tilde{\chi}_1^+ \rightarrow
  \tilde{\chi}_1^0 \mu^+ \nu_{\mu}$ (\cite{kkk}); (b) $\tilde{\chi}_1^+
  \rightarrow \nu_{\ell} \mu^+ \nu$ (\cite{krsz}).}
\end{figure}

In~\cite{kkk}, the $\tilde{t}$ is produced in collisions of positrons
with valence $d$-quarks in the proton, \ie $k=1$ in equation~\ref{lagr}
(down-stop scenario), whereas in~\cite{krsz} the case $e^+s\rightarrow
\tilde{t}$ ($k=2$, strange-stop scenario) is considered in addition.
The papers also differ in the assumed squark decay chains, shown in
figure~\ref{decays}, that give rise to the characteristic features of
the muon events.

For a stop mass in the range 100 -- 200~${\rm GeV}$ and with
$\lambda'_{131}$ large enough for stop production to be relevant, there
is a wide range of parameters where the decay $\tilde{t}\rightarrow
b\tilde{\chi}^+_1$ dominates over other decay modes.  Then the decay
chain shown in figure~\ref{decays}a, $\tilde{t} \rightarrow
b\tilde{\chi}_1^+$, $\tilde{\chi}_1^+ \rightarrow
\mu^+\nu\tilde{\chi}_1^0$ generates an isolated muon and a $b$ quark jet
at large transverse momenta. Since the muon originates from the virtual
$W^+$, similar events with isolated positrons should be observed.
Moreover, to account for the topology of the events with large missing
$P_T$, the neutralino $\tilde{\chi}_1^0$ must be assumed very long-lived
($\Gamma_{\tilde{\chi}^0_1} \lesssim 10^{-7}$~${\rm eV}$), so that it
may escape detection despite the presence of $R_{p}$-conserving decay
channels. Since parameters are not easily arranged that give rise to
such a long lifetime, alternative decay channels have been considered
in~\cite{krsz}.

If trilinear lepton couplings $\lambda LL{E}^c$ are also present in
supersymmetric theories with sleptons in the mass range of 100 to
200~${\rm GeV}$, another possibility for stop and subsequent chargino
decays is open, as shown in figure~\ref{decays}b.  The chargino may
decay into a neutrino and a slepton, followed by the $R_p$-violating
slepton decay to a positively charged muon and a neutrino.  Such a chain
can account for the observed final state, \ie a jet, a single positively
charged muon and missing transverse momentum.

In the case of stop production in $e^+d$ collisions (down stop
scenario), the chargino must be heavy with $m_{\tilde{\chi}^+_1}\approx
180$ -- $190$~${\rm GeV}$ to account for the required balance of
$R_p$-conserving and violating stop decay modes implied by the
low-energy, HERA and Tevatron data. By contrast, for the strange stop
scenario $e^+s \rightarrow \tilde{t}$ with a larger value of
$\lambda'_{132}$ than $\lambda'_{131}$, one finds a solution for lighter
chargino masses $m_{\tilde{\chi}^+_1}\sim 100$ -- 140~${\rm GeV}$.

Assuming a given value for $m_{\tilde{t}}$, the mass
$m_{\tilde{\chi}_1^+}$ recoiling against the hadronic $b$ jet can be
estimated from the calculated 4-momentum of the top squark and the
measured 4-momentum of the $b$ jet: $m^2_{\tilde{\chi}_1^+} = \left(
  p_{\tilde{t}} - p_b \right)^2$. The recoil masses must cluster for the
observed events; if not, two-body decays of the stop resonance are not
the origin of the events, or more than one stop is produced. It is
amusing to observe that if both stops with masses
$m_{\tilde{t}_1}=200$~${\rm GeV}$ and $m_{\tilde{t}_2}=230$~${\rm GeV}$
are responsible for the H1 events, the estimated recoiling mass
$m_{\tilde{\chi}_1^+}$ falls in the range 130 -- 140~${\rm GeV}$,
compatible with the strange stop scenario.
 
The branching ratio for the chargino decay $\tilde{\chi}_1^+ \rightarrow
\nu_{\ell} \tilde{\ell}^+$ can be expected to be close to 1/6. The
subsequent decay $\tilde{\ell}^+ \rightarrow \mu^+\nu$ has to compete
with other $R_{p}$-violating and also with $R_{p}$-conserving decay
modes.  The semi-quantitative discussion performed in \cite{krsz}
suggests that a decay chain $\tilde{t} \rightarrow b \tilde{\chi}_1^+
\rightarrow b \nu \tilde{\ell}^+ \rightarrow b\nu\nu\mu^+$, leading to
the observed topology, could be realized in supersymmetric theories with
$R_{p}$-breaking couplings. However, a large number of other final
states with rather complex topologies should be observed at HERA in
$e^+p$ collisions generated by the mixed $R_{p}$-conserving and
violating decay modes. Single and multi-lepton states associated with
one or more jets and, in most cases, missing transverse momentum due to
escaping neutrinos can be expected,
\begin{equation}
\begin{array}{l}
e^+p \rightarrow \tilde{t} \rightarrow \tilde{\chi}_1^+ b
\rightarrow \ell^+{\rm j}\nu\nu, ~ \ell^+ \ell^+\ell^- {\rm j}, ~
\ell^+ {\rm jjj}, ~ {\rm jjj}\nu
\\[1ex]
e^+p \rightarrow \tilde{c} \rightarrow \tilde{\chi}_1^0 c
\rightarrow  \ell^+ \ell^- {\rm j}\nu, ~
\ell^{\pm} {\rm jjj}, ~ {\rm jjj}\nu
\end{array}
\end{equation}
where $\ell$, $\nu$, j generically denote charged leptons, neutrinos and
jets.  However, not all combinations are possible in principle. For
example single negatively charged lepton events can be accompanied by
jets but not by neutrinos. Kinematical constraints imposed by the fixed
masses of the intermediate supersymmetric particles can be exploited to
check whether such hypothetical decay chains are realized or not.

From the above discussion it is clear that isolated $\mu^+$ events in
$e^+p$ scattering can occur in supersymmetric scenarios with
$R_p$-violating interactions. The presence of both $\lambda'
{L}{Q}{D}^c$ and $\lambda {L}{L}{E}^c$ terms in the superpotential
provides a large variety of mechanisms.  If true, a wealth of other
interesting phenomena could be observed, not only at HERA.

\section{Summary and Conclusions}

The ZEUS results, along with the kinematic analysis of the events
themselves, have shown that $W$ production alone is unlikely to account
for all the H1 high $P_{T}$ lepton events.  Moreover, it is likely that
events of a similar topology to those observed by H1 would have been
found by ZEUS in a similar high $P_{T}$ track based search. It is
intriguing that high $P_{T}$ lepton events of the kind observed by H1
can naturally arise in certain $R_{p}$-violating SUSY scenarios.
Nevertheless, only more data will allow the source of the events to be
finally established.

\ack JK has been partially supported by the Polish Committee for
Scientific Research Grant 2 P03B 030 14.  TM and DSW have been assisted
by the British Council, Collaborative Research Project TOK/880/11/15.


\section{References}

\begin{thebibliography}{99}
  
\bibitem{h1_iso_lepton} Adloff C {\em et al.} (H1 Collaboration) 1998
  {\it Eur.\ Phys.\ J.\/ } {\bf 5} 575
  
\bibitem{zeus_w} ZEUS Collaboration 1998 {\it Measurement of the $W$
    Production Cross Section in $e^{+}p$ Collisions at HERA} paper 756
  contributed to ICHEP98 Vancouver
  
\bibitem{BVZ_w} Baur U, Vermaseren J A M and Zeppenfeld D 1992 \NP {\bf
    375} 3
  
\bibitem{spira} P.Nason, M.Spira, R.R\"uckl (unpublished) ; M.Spira,
  private communication.
  
\bibitem{plenary_proc} Waters D S 1998 {\it Exotic and Rare Processes at
    HERA} Proceedings of the 3rd UK Phenomenology Workshop on HERA
  Physics, September 1998, to be published in Journal of Physics G.
  
\bibitem{kkk} Kon T, Kobayashi T and Kitamura S 1996 \PL {\bf 376} 227
  
\bibitem{krsz} Kalinowski J, R\"uckl R, Spiesberger H and Zerwas P M
  1997 {\it Speculations on SUSY Mechanisms of Isolated $\mu^+$
    Production in $e^+p$ Collisions} DESY Internal Note (unpublished)

\bibitem{barbier} Barbier R \etal 1998 {\it hep-ph/9810232}

\end{thebibliography}
\end{document}